\documentclass[10pt]{iopart}

\usepackage{graphicx}
\usepackage{iopams}

\begin{document}

\title{Two-dimensional extended Hubbard model at half-filling}%

\author{A Sherman}%

\address{Institute of Physics, University of Tartu, W. Ostwaldi Str 1, 50411 Tartu, Estonia}

\ead{alexeisherman@gmail.com}

\begin{abstract}
We consider the extended Hubbard model on a two-dimensional square lattice at half-filling. The model is investigated using the strong coupling diagram technique. We sum infinite series of ladder diagrams allowing for full-scale charge and spin fluctuations and the actual short-range antiferromagnetic order for nonzero temperatures. In agreement with earlier results, we find the first-order phase transition in the charge subsystem occurring at $v=v_c\gtrsim U/4$ with $v$ and $U$ the intersite and on-site Coulomb repulsion constants. The transition reveals itself in an abrupt sign change of a sharp maximum in the zero-frequency charge susceptibility at the corner of the Brillouin. States arising at the transition have alternating deviations of electron occupations from the mean value on neighboring sites. Due to fluctuations, these alternating occupation deviations have short-range order. For the considered parameters, such behavior is found for $U\lesssim5t$ with $t$ the hopping constant. For the insulating case $U\gtrsim 6t$, in which the transition is not observed, we find a continuous growth of the Mott gap with $v$. The evolution of the electron density of states with increasing $v$ is also considered.
\end{abstract}

\vspace{2pc}


\maketitle


\section{Introduction}
The extended Hubbard model (EHM) is a generalization of the Hubbard model, which allows one to study the influence of the non-local Coulomb interaction on the properties of the strongly correlated electron system. Along with the on-site repulsion and kinetic energy, the EHM Hamiltonian contains the term describing an interaction of electrons on neighboring sites. The incomplete screening of this coupling was found in several low-dimensional crystals such as graphene \cite{Kotov}, Bechgaard salts \cite{Pariser}, and polymers \cite{Friend}. A sizable non-local interaction was expected in cuprate perovskites \cite{Hozoi} influencing the charge separation \cite{Citro}. Early studies of EHM were carried out using Monte-Carlo simulations \cite{Hirsch,Lin,Zhang}, exact solutions for small clusters \cite{Fourcade,Bosch}, and mean-field approximations \cite{Yan,Dagotto}. These works demonstrated that the electron repulsion on neighboring sites leads to the phase transition in the charge subsystem occurring at $v_c\approx U/z$. Here $U$ and $v$ are Coulomb interaction constants for electrons on the same and neighboring sites, and $z$ is the coordination number. The transition was connected with the appearance of states having alternating deviations of electron occupations from the mean value on neighboring sites (SAOs for short). Since the statistical ensemble contains states with both deviation signs, the mean site occupation remains uniform throughout the crystal. At half-filling, such deviations decrease the site spin. Therefore, the transition suppresses the antiferromagnetic ordering (AFO) of electron spins. Later on, the EHM was investigated using the extended dynamic mean-field theory (DMFT) \cite{Sun}, its diagrammatic extensions \cite{Ayral,Loon}, cluster generalizations of the DMFT \cite{Paki}, variational cluster approximation \cite{Aichhorn}, and the two-particle self-consistent approach \cite{Davoudi}. In these works, the transition was shown to be of the first order, and phase diagrams were obtained. For low temperatures $T$ and $U$ smaller than the critical value of the Mott transition, the metallic and SAO phases were separated by a nearly straight line in the $U$-$v$ plane \cite{Sun,Ayral,Loon,Paki}. The line was almost parallel to the mean-field borderline $v=U/z$. In the $T$-$v$ plane, the boundary was also close to linear, and it had a positive slope \cite{Paki}.

This work uses the strong coupling diagram technique (SCDT) \cite{Vladimir,Metzner,Pairault,Sherman18} to investigate the half-filled EHM on a two-dimensional (2D) square lattice. In contrast to the above works, this approach allows us to properly account for full-scale charge and spin fluctuations and the short-range AFO in the crystal at finite temperatures. The adequate description of spin excitations is significant for obtaining reliable electron spectra. Another advantage of the used approach is its applicability for any non-local interaction between electrons if it is smaller than the on-site coupling.

We consider the ranges of parameters $2t\leq U\leq 8t$, $v\lesssim U/2$ and $0.1t\lesssim T\ll U$ with $t$ the hopping constant. We sum infinite series of ladder diagrams to derive a closed set of SCDT equations. Self-consistent solutions of these equations are obtained by iteration. For $U\lesssim5t$, we find the phase transition in the charge subsystem at $v_c\gtrsim U/4$. The transition reveals itself in an abrupt sign change of the zero-frequency charge susceptibility $\chi^{\rm ch}({\bf Q},0)$ at the corner of the Brillouin zone ${\bf Q}$. At $v$ near $v_c$, the susceptibility peaks sharply at this momentum. In contrast to the mentioned mean-field approaches, $\chi^{\rm ch}({\bf Q},0)$ remains finite at $v=v_c$, which indicates a short-range ordering of alternating occupations. The temperature dependence of the susceptibility suggests that this type of order stems from the fluctuations taken into account in this work. The transition is of the first order -- for $v$ near $v_c$, two solutions of the considered set of equations coexist. The correlation length of the short-range AFO $\xi$ decreases monotonously with increasing $v$. This length remains larger than the intersite distance up to $v=v_c$. For $U\lesssim5t$ and nonzero temperature, the electron spectrum is metallic. The density of electron states (DOS) near the Fermi level decreases monotonously as $v$ approaches $v_c$. For larger on-site repulsions, $U\gtrsim6t$, when the DOS contains the Mott gap, the transition is not observed in the considered ranges of intersite repulsions and temperatures. In this case, the growth of $v$ leads to a monotonous increase in the width of the Mott gap.

The paper is organized as follows: The model Hamiltonian, a brief discussion of the SCDT, and the main formulas are given in the next section. Manifestations of the transition in determinants of the Bethe-Salpeter equations (BSE), charge susceptibility, and shapes of DOSs are considered in Sect.~3. The last section is devoted to concluding remarks.

\section{Model and SCDT}
The EHM Hamiltonian reads
\begin{eqnarray}\label{Hamiltonian}
H&=&\sum_{\bf ll'\sigma}t_{\bf ll'}a^\dagger_{\bf l'\sigma}a_{\bf l\sigma}+\frac{U}{2}\sum_{\bf l\sigma}n_{\bf l\sigma}n_{\bf l,-\sigma}\nonumber\\
&&+\frac{1}{2}\sum_{\bf ll'}v_{\bf ll'}\big(n_{\bf l'}-\bar{n}\big)\big(n_{\bf l}-\bar{n}\big)-\mu\sum_{\bf l}n_{\bf l},
\end{eqnarray}
where {\bf l} and ${\bf l'}$ are site vectors of a 2D square lattice, $\sigma=\pm1$ is the spin projection, $a^\dagger_{\bf l\sigma}$ and $a_{\bf l\sigma}$ are electron creation and annihilation operators, $t_{\bf ll'}$, $U$, and $v_{\bf ll'}$ are constants of hopping, on-site and intersite Coulomb repulsions, respectively, $n_{\bf l\sigma}=a^\dagger_{\bf l\sigma}a_{\bf l\sigma}$ and $n_{\bf l}=\sum_\sigma n_{\bf l\sigma}$ are site occupation numbers, $\bar{n}=\langle n_{\bf l}\rangle$ is the occupation mean value with the angle brackets denoting the statistical averaging, and $\mu$ is the chemical potential. In this work, $t_{\bf ll'}$ and $v_{\bf ll'}$ are supposed to be nonzero for neighboring sites only,
\begin{equation*}
t_{\bf ll'}=-t\sum_{\bf a}\delta_{\bf l',l+a},\quad v_{\bf ll'}=v\sum_{\bf a}\delta_{\bf l',l+a},
\end{equation*}
where {\bf a} are four vectors connecting neighboring sites. Below, we consider the case of half-filling, $\bar{n}=1$, which for the Hamiltonian~(\ref{Hamiltonian}) takes place at $\mu=U/2$.

For calculating Green's functions, we use the SCDT \cite{Vladimir,Metzner,Pairault,Sherman18}. Supposing that the on-site Coulomb repulsion is the largest energy parameter, in this approach, the local part of the Hamiltonian is considered as an unperturbed operator $H_0$, and correlators are calculated using series expansions in powers of nonlocal terms $H_i$ (see, e.g., \cite{Sherman18,Sherman20a,Sherman20b}). In the present case, the on-site Coulomb interaction and the chemical-potential term of the Hamiltonian (\ref{Hamiltonian}) form $H_0$, while other parts $H_i$. Terms of the SCDT series are products of the hopping and intersite Coulomb interaction constants and on-site cumulants \cite{Kubo} of electron creation and annihilation operators. We consider terms with cumulants of the first and second orders only. This approximation was enough to obtain quantitatively correct results in the Hubbard model \cite{Sherman18,Sherman21}. These cumulants are con\-s\-truc\-ted from a pair and two pairs of creation and annihilation operators,
\begin{eqnarray*}
&&C^{(1)}(\tau',\tau)=\big\langle{\cal T}\bar{a}_{{\bf l}\sigma}(\tau')a_{{\bf l}\sigma}(\tau)\big\rangle_0,\\
&&C^{(2)}(\tau_1,\sigma_1;\tau_2,\sigma_2;\tau_3,\sigma_3;\tau_4,\sigma_4)\\
&&\quad=\big\langle{\cal T}\bar{a}_{{\bf l}\sigma_1}(\tau_1)a_{{\bf l}\sigma_2}(\tau_2) \bar{a}_{{\bf l}\sigma_3}(\tau_3)a_{{\bf l}\sigma_4}(\tau_4)\big\rangle_0\\
&&\quad\quad-\big\langle{\cal T}\bar{a}_{{\bf l}\sigma_1}(\tau_1)a_{{\bf l}\sigma_2}(\tau_2)\big\rangle_0\big\langle{\cal T}\bar{a}_{{\bf l}\sigma_3}(\tau_3)a_{{\bf l}\sigma_4}(\tau_4)\big\rangle_0\\
&&\quad\quad+\big\langle{\cal T}\bar{a}_{{\bf l}\sigma_1}(\tau_1)a_{{\bf l}\sigma_4}(\tau_4)\big\rangle_0\big\langle{\cal T}\bar{a}_{{\bf l}\sigma_3}(\tau_3)a_{{\bf l}\sigma_2}(\tau_2)\big\rangle_0.
\end{eqnarray*}
The subscript 0 at angle brackets indicates that time dependencies of operators and the statistical averaging are determined by the site Hamiltonian
\begin{equation*}
H_{\bf l}=\sum_\sigma\big[(U/2)n_{\bf l\sigma}n_{\bf l,-\sigma}-\mu n_{\bf l\sigma}\big].
\end{equation*}
The sum of the site Hamiltonians forms $H_0$. The symbol ${\cal T}$ is the chronological ope\-ra\-tor.

The terms of the SCDT series expansion can be visualized by depicting $t_{\bf ll'}$ as directed lines, $v_{\bf ll'}$ as crosses, and cumulants as circles. The number of lines outgoing from and incoming to the circle indicates the number of electron operators in the cumulant. As for the weak coupling diagram technique \cite{Abrikosov}, the linked-cluster theorem is valid and partial summations are allowed in the SCDT. The notion of the one-particle irreducible diagram can also be introduced in this diagram technique. It is a two-leg diagram, which cannot be divided into two disconnected parts by cutting a hopping line $t_{\bf ll'}$. If we denote the sum of all such diagrams -- the irreducible part -- by the symbol $K$, the Fourier transform of the electron Green's function $G({\bf l'\tau',l\tau})=\langle{\cal T}\bar{a}_{\bf l'\sigma}(\tau')a_{\bf l\sigma}(\tau)\rangle$ can be written as
\begin{equation}\label{Larkin}
G({\bf k},j)=\big\{\big[K({\bf k},j)\big]^{-1}-t_{\bf k}\big\}^{-1}.
\end{equation}
Here ${\bf k}$ is the 2D wave vector and the integer $j$ defines the Matsubara frequency $\omega_j=(2j-1)\pi T$.

\begin{figure}[t]
\centerline{\resizebox{0.8\columnwidth}{!}{\includegraphics{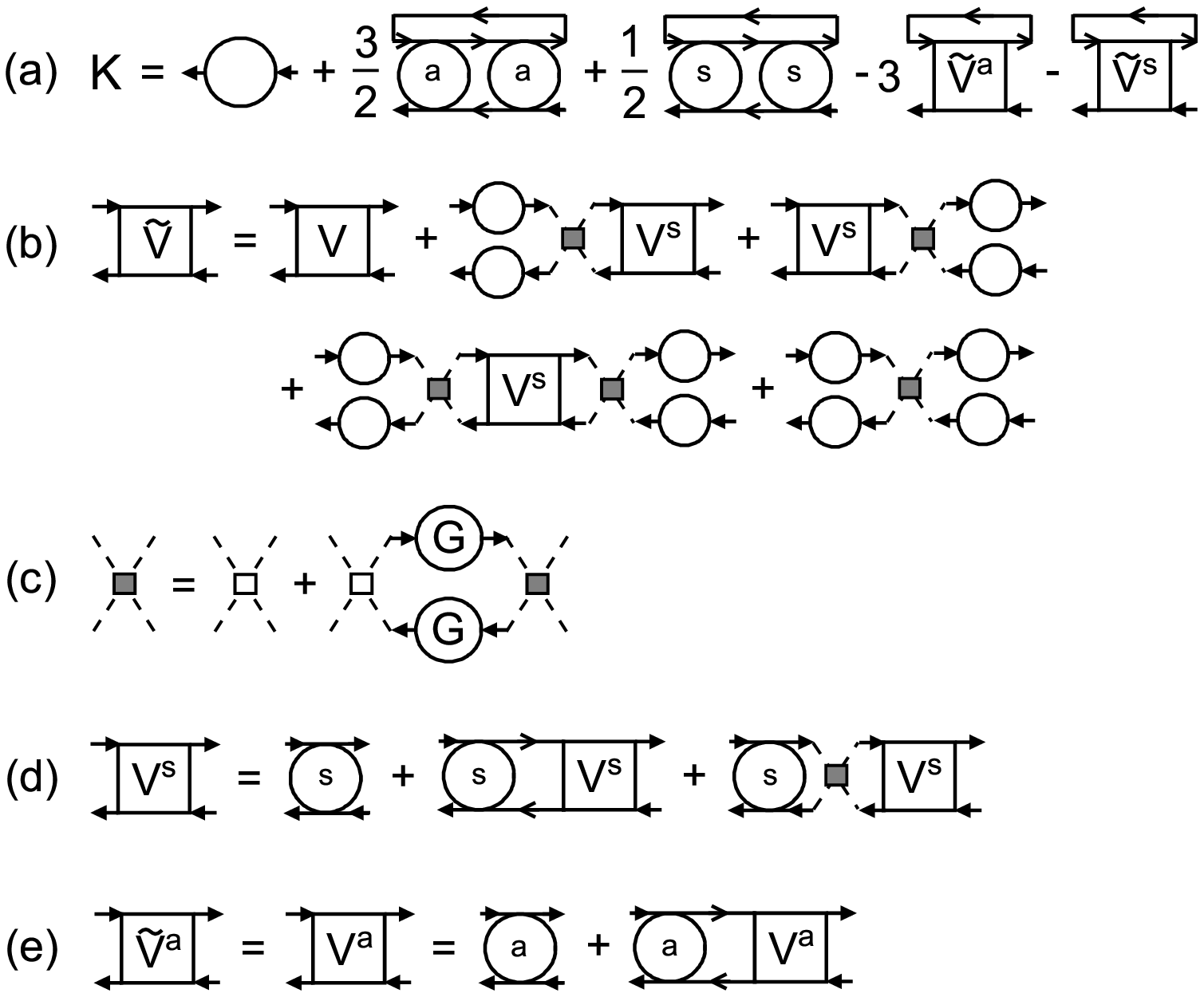}}}
\caption{The diagrammatic representation of main equations. Open circles with one incoming and one outgoing arrows are first-order cumulants, circles with letters s and a are symmetrized and antisymmetrized second-order cumulants, circles with the letter G are Green's functions, arrowed solid lines are renormalized hopping $\theta({\bf k},j)$, Eq.~(\ref{theta}), dashed crosses with open small squares are bare intersite Coulomb vertices $v_{\bf k}$, crosses with greyed squares are renormalized vertices $\varphi({\bf k},j)$, Eq.~(\ref{phi}), squares with letters $\widetilde{V}^s$, $V^s$, $\widetilde{V}^a$ and $V^a$ are infinite sums of ladder diagrams in part (b) symmetrized and antisymmetrized over spin indices, Eqs.~(\ref{Vts})--(\ref{Va}).} \label{Fig1}
\end{figure}
Diagrams taken into account in the present calculations are shown in Fig.~\ref{Fig1}. Here short arrows entering and leaving cumulants and vertices shown by squares are their endpoints, the solid arrowed lines connecting these endpoints are the renormalized hopping
\begin{equation}\label{theta}
\theta({\bf k},j)=t_{\bf k}+t_{\bf k}^2G({\bf k},j),
\end{equation}
the dashed cross with an open square in the center is the bare intersite Coulomb repulsion $v_{\bf k}$, and the similar cross with a greyed square is the renormalized Coulomb interaction, Fig.~\ref{Fig1}(c),
\begin{equation}\label{phi}
\varphi({\bf k},\nu)=\frac{v_{\bf k}}{1-2v_{\bf k}TN^{-1}\sum_{{\bf q}j}G({\bf q},j)G({\bf k+q},\nu+j)}.
\end{equation}
Here $N$ is the number of sites. Algebraically diagrams in Fig.~\ref{Fig1}(a) read
\begin{eqnarray}\label{K}
K({\bf k},j)&=&C^{(1)}(j)+\frac{T^2}{4N}\sum_{{\bf k'}j'\nu}\theta({\bf k'},j') {\cal T}_{\bf k-k'}(j+\nu,j'+\nu)\nonumber\\
&&\times\big[3C^{(a)}(j,j+\nu,j'+\nu,j')C^{(a)}(j+\nu,j,j',j'+\nu)\nonumber\\
&&\quad+C^{(s)}(j,j+\nu,j'+\nu,j')C^{(s)}(j+\nu,j,j',j'+\nu)\big]\nonumber\\
&&-\frac{T}{2N}\sum_{{\bf k'}j'}\theta({\bf k'},j')\big[3\widetilde{V}_{\bf k-k'}^{(a)}(j,j,j',j')\nonumber\\
&&\quad+\widetilde{V}_{\bf k-k'}^{(s)}(j,j,j',j')\big],
\end{eqnarray}
where ${\cal T}_{\bf k}(j,j')=N^{-1}\sum_{\bf k'}\theta({\bf k+k'},j)\theta({\bf k'},j')$, $C^{(s)}$ and $C^{(a)}$ are second-order cumulants symmetrized and antisymmetrized over spin indices,
\begin{eqnarray}\label{Cas}
&&C^{(s)}(j+\nu,j,j',j'+\nu)=\sum_{\sigma'}C^{(2)}(j+\nu,\sigma';j,\sigma;j',\sigma; j'+\nu,\sigma')\nonumber\\[-1ex]
&&\\[-1ex]
&&C^{(a)}(j+\nu,j,j',j'+\nu)=\sum_{\sigma'}\sigma\sigma'C^{(2)}(j+\nu,\sigma';j, \sigma; j',\sigma;j'+\nu,\sigma'),\nonumber
\end{eqnarray}
quantities $\widetilde{V}^{(s)}$ and $\widetilde{V}^{(a)}$ are results of the analogous symmetrization and an\-ti\-sym\-met\-ri\-za\-ti\-on of the infinite sum of ladder diagrams $\widetilde{V}$ in Fig.~\ref{Fig1}(b). Two-particle irreducible vertices in this sum are the renormalized Coulomb vertices (\ref{phi}) and second-order cumulants. If the former vertices allow for the intersite Coulomb interaction, the latters describe the on-site coupling. All possible sequences of these vertices are taken into account,
\begin{eqnarray}
&&\widetilde{V}^{(s)}_{\bf k}(j+\nu,j,j',j'+\nu)=V^{(s)}_{\bf k}(j+\nu,j,j',j'+\nu)\nonumber\\
&&\quad+C^{(1)}(j+\nu)C^{(1)}(j'+\nu)\varphi({\bf k},j-j')\nonumber\\
&&\quad\quad\times T\sum_{\nu'}V^{(s)}_{\bf k}(j+\nu',j,j',j'+\nu')\nonumber\\
&&\quad+C^{(1)}(j)C^{(1)}(j')\varphi({\bf k},j-j')\nonumber\\
&&\quad\quad\times T\sum_{\nu'}V^{(s)}_{\bf k}(j+\nu,j+\nu',j'+\nu',j'+\nu)\nonumber\\
&&\quad+2C^{(1)}(j+\nu)C^{(1)}(j'+\nu)C^{(1)}(j)C^{(1)}(j')\varphi^2({\bf k},j-j')\nonumber\\
&&\quad\quad\times T^2\sum_{\nu'\nu''}V^{(s)}_{\bf k}(j+\nu',j+\nu'',j'+\nu'',j'+\nu') \nonumber\\
&&\quad+C^{(1)}(j+\nu)C^{(1)}(j'+\nu)C^{(1)}(j)C^{(1)}(j')\varphi({\bf k},j-j'), \label{Vts}\\[1ex]
&&\widetilde{V}^{(a)}_{\bf k}(j+\nu,j,j',j'+\nu)=V^{(a)}_{\bf k}(j+\nu,j,j',j'+\nu).\label{Vta}
\end{eqnarray}
Here $V^{(s)}$ and $V^{(a)}$ are parts of these ladders, which start and finish with the second-order cumulants, Fig.~\ref{Fig1}(d) and (e),
\begin{eqnarray}
&&V^{(s)}_{\bf k}(j+\nu,j,j',j'+\nu)=C^{(s)}(j+\nu,j,j',j'+\nu)\nonumber\\
&&\quad+T\sum_{\nu'} C^{(s)}(j+\nu,j+\nu',j'+\nu',j'+\nu){\cal T}_{\bf k}(j+\nu',j'+\nu') \nonumber\\
&&\quad\quad\times V^{(s)}_{\bf k}(j+\nu',j,j',j'+\nu')\nonumber\\
&&\quad+2T^2\sum_{\nu'\nu''}C^{(s)}(j+\nu,j+\nu',j'+\nu',j'+\nu)\varphi({\bf k},j-j') \nonumber\\
&&\quad\quad\times V^{(s)}_{\bf k}(j+\nu'',j,j',j'+\nu'')\label{Vs}\\[1ex]
&&V^{(a)}_{\bf k}(j+\nu,j,j',j'+\nu)=C^{(a)}(j+\nu,j,j',j'+\nu)\nonumber\\
&&\quad+T\sum_{\nu'} C^{(a)}(j+\nu,j+\nu',j'+\nu',j'+\nu){\cal T}_{\bf k}(j+\nu',j'+\nu') \nonumber\\
&&\quad\quad\times V^{(a)}_{\bf k}(j+\nu',j,j',j'+\nu').\label{Va}
\end{eqnarray}

Reducible vertices $\widetilde{V}^{(s)}$ and $\widetilde{V}^{(a)}=V^{(a)}$ describe charge and spin fluctuations and define respective susceptibilities $\chi^{\rm ch}({\bf l'}\tau',{\bf l}\tau)=\frac{1}{2}\langle{\cal T}(n_{\bf l'}(\tau')-\bar{n})(n_{\bf l}(\tau)-\bar{n})\rangle$ and $\chi^{\rm sp}({\bf l'}\tau',{\bf l}\tau)=\langle{\cal T}\bar{a}_{\bf l'\sigma}(\tau')a_{\bf l',-\sigma}(\tau')\bar{a}_{\bf l,-\sigma}(\tau)a_{\bf l\sigma}(\tau)\rangle$,
\begin{eqnarray}
&&\chi^{\rm ch}({\bf k},\nu)=-\frac{T}{N}\sum_{{\bf q}j}G({\bf k+q},\nu+j)G({\bf k},j) \nonumber\\
&&\quad-T^2\sum_{jj'}F_{\bf k}(j,\nu+j)F_{\bf k}(j',\nu+j')\widetilde{V}^{(s)}_{\bf k}(\nu+j,\nu+j' ,j',j),\label{chich}\\
&&\chi^{\rm sp}({\bf k},\nu)=-\frac{T}{N}\sum_{{\bf q}j}G({\bf k+q},\nu+j)G({\bf k},j) \nonumber\\
&&\quad-T^2\sum_{jj'}F_{\bf k}(j,\nu+j)F_{\bf k}(j',\nu+j')V^{(a)}_{\bf k}(\nu+j,\nu+j' ,j',j),\label{chisp}
\end{eqnarray}
where $F_{\bf k}(j,j')=N^{-1}\sum_{\bf q}\Pi({\bf q},j)\Pi({\bf k+q},j')$ and $\Pi({\bf k},j)=1+t_{\bf k}G({\bf k},j)$.

Notice that the inclusion of the intersite Coulomb interaction does not directly modify Eqs.~(\ref{Vta}) and (\ref{Va}) for the spin vertex $\widetilde{V}^{(a)}$. The latter equation looks similar to the respective formula in the Hubbard model \cite{Sherman18}. In this equation, the influence of the intersite repulsion is indirect, through the modification of electron Green's functions entering into the quantity ${\cal T}_{\bf k}(j,j')$. Comparing with the Hubbard model, we see that main changes occurred in Eqs.~(\ref{Vts}) and (\ref{Vs}) for the charge vertex $\widetilde{V}^{(s)}$.

For calculations, the above formulas have to be supplemented by expressions for cumulants. They can be found in \cite{Vladimir,Metzner,Pairault,Sherman18}. These expressions can be significantly simplified in the case
\begin{equation}\label{condition}
T\ll\mu,\quad T\ll U-\mu.
\end{equation}
For $U\gg T$, this range of chemical potentials contains relevant cases of half-filling, $\mu=U/2$, and moderate doping. In this range, cumulants read
\begin{eqnarray}\label{cumulants}
&&C^{(1)}(j)=\frac{1}{2}\big[g_1(j)+g_2(j)\big],\nonumber \\
&&C^{(2)}(j+\nu,\sigma;j,\sigma';j',\sigma';j'+\nu,\sigma)=\frac{1}{4T}\big[\delta_{jj'}\big(1-2 \delta_{\sigma\sigma'}\big)\nonumber\\
&&\quad+\delta_{\nu0}\big(2-\delta_{\sigma\sigma'}\big)\big]a_1(j'+\nu)a_1(j)-\frac{1}{2} \delta_{\sigma,-\sigma'}\big[a_1(j'+\nu)a_2(j,j')\\
&&\quad+a_2(j'+\nu,j+\nu)a_1(j)+a_3(j'+\nu,j+\nu)a_4(j,j')\nonumber\\
&&\quad+a_4(j'+\nu,j+\nu)a_3(j,j')\big],\nonumber
\end{eqnarray}
where
\begin{eqnarray*}
&&g_1(j)=({\rm i}\omega_j+\mu)^{-1},\quad g_2(j)=({\rm i}\omega_j+\mu-U)^{-1}, \\
&&a_1(j)=g_1(j)-g_2(j),\quad a_2(j,j')=g_1(j)g_1(j'),\\
&&a_3(j,j')=g_2(j)-g_1(j'),\quad a_4(j,j')=a_1(j)g_2(j').
\end{eqnarray*}

With these expressions, vertices $V^{(s)}$ and $V^{(a)}$ acquire the form
\begin{eqnarray}
&&V_{\bf k}^{(s)}(j+\nu,j,j',j'+\nu)=\frac{1}{2}f_{\bf k}^{(2)}(j+\nu,j'+\nu) \nonumber\\
&&\quad\times\big\{2C^{(s)}(j+\nu,j,j',j'+\nu)+4a_{\bf k}(j'+\nu,j+\nu)\nonumber\\
&&\quad\times[C'(j,j')+z'({\bf k},j,j')]-a_2(j'+\nu,j+\nu)z_1({\bf k},j,j') \nonumber\\
&&\quad-a_1(j'+\nu)z_2({\bf k},j,j')-a_4(j'+\nu,j+\nu)z_3({\bf k},j,j')\nonumber\\
&&\quad-a_3(j'+\nu,j+\nu)z_4({\bf k},j,j')\big\}\label{Vs2}\\
&&V_{\bf k}^{(a)}(j+\nu,j,j',j'+\nu)=\frac{1}{2}f_{\bf k}^{(1)}(j+\nu,j'+\nu) \nonumber\\
&&\quad\times\big\{2C^{(a)}(j+\nu,j,j',j'+\nu)+\big[a_2(j'+\nu,j+\nu)\nonumber\\
&&\quad-T^{-1}\delta_{jj'}a_1(j'+\nu)\big]y_1({\bf k},j,j')+a_1(j'+\nu)y_2({\bf k},j,j') \nonumber\\
&&\quad+a_4(j'+\nu,j+\nu)y_3({\bf k},j,j')+a_3(j'+\nu,j+\nu)y_4({\bf k},j,j')\big\}\label{Va2}.
\end{eqnarray}
The BSE (\ref{Vs}) and (\ref{Va}) are transformed into two small systems of linear equations. Each system has four equations with four unknowns $z_i({\bf k},j,j')$ or $y_i({\bf k},j,j')$, $i=1,\ldots 4$, for fixed ${\bf k}$, $j$ and $j'$,
\begin{eqnarray}
&&z_i({\bf k},j,j')=d_i({\bf k},j,j')-e_{i2}({\bf k},j-j')z_1({\bf k},j,j')\nonumber \\
&&\quad-\big[e_{i1}({\bf k},j-j')-p_i({\bf k},j-j')\big]z_2({\bf k},j,j')\nonumber\\
&&\quad-\big[e_{i4}({\bf k},j-j')-\big({\rm i}\omega_j-{\rm i}\omega_{j'}-U\big)^{-1}p_i({\bf k},j-j')\big]z_3({\bf k},j,j')\nonumber\\
&&\quad-\big[e_{i3}({\bf k},j-j')+p_i({\bf k},j-j')\big]z_4({\bf k},j,j'),\label{zi}\\
&&y_i({\bf k},j,j')=b_i({\bf k},j,j')+\big[c_{i2}({\bf k},j-j')-T^{-1}\delta_{jj'}c_{i1}({\bf k},j-j')\big]\nonumber \\
&&\quad\times y_1({\bf k},j,j')+c_{i1}({\bf k},j-j')y_2({\bf k},j,j')+e_{i4}({\bf k},j-j')y_3({\bf k},j,j')\nonumber\\
&&\quad+c_{i3}({\bf k},j-j')y_4({\bf k},j,j').\label{yi}
\end{eqnarray}
Hence the BSE equations~(\ref{Vs}) and (\ref{Va}) can be exactly solved. In the above relations
\begin{eqnarray*}
&&f^{(1)}_{\bf k}(j,j')=\bigg[1+\frac{1}{4}a_1(j)a_1(j'){\cal T}_{\bf k}(j,j')\bigg],\\
&&f^{(2)}_{\bf k}(j,j')=\bigg[1-\frac{3}{4}a_1(j)a_1(j'){\cal T}_{\bf k}(j,j')\bigg],\\
&&C'(j,j')=T\sum_\nu C^{(s)}(j+\nu,j,j',j'+\nu)\\
&&\quad=\frac{1}{4}\bigg[a_2(j,j')-a_4(j,j')+\frac{a_3(j,j')}{{\rm i}\omega_j-{\rm i}\omega_{j'}-U}\bigg],\\
&&C''(j-j')=T\sum_\nu C'(j+\nu,j'+\nu)=\frac{U/2}{(\omega_j-\omega_{j'})^2+U^2},\\
&&a_{\bf k}(j,j')=\frac{\varphi({\bf k},j-j')C'(j,j')}{1-2C''(j-j')\varphi({\bf k},j-j')},\\
&&z'({\bf k},j,j')=\frac{1}{4}\bigg[z_2({\bf k},j,j')-z_4({\bf k},j,j')+\frac{z_3({\bf k},j,j')}{{\rm i}\omega_j-{\rm i}\omega_{j'}-U}\bigg],\\
&&e_{ii'}({\bf k},\nu)=\frac{T}{2}\sum_j a_i(\nu+j,j)a_{i'}(j,\nu+j){\cal T}_{\bf k}(\nu+j,j) f^{(2)}_{\bf k}(\nu+j,j),\\
&&c_{ii'}({\bf k},\nu)=\frac{T}{2}\sum_j a_i(\nu+j,j)a_{i'}(j,\nu+j){\cal T}_{\bf k}(\nu+j,j) f^{(1)}_{\bf k}(\nu+j,j),\\
&&p_i({\bf k},\nu)=\frac{\varphi({\bf k},\nu)}{4[1-2C''(\nu)\varphi({\bf k},\nu)]}\\
&&\quad\times\bigg[e_{i2}({\bf k},\nu)-e_{i4}({\bf k},\nu)-\frac{e_{i3}({\bf k},\nu)}{{\rm i}\omega_{j'}-{\rm i}\omega_{j}-U}\bigg],\\
&&d_i({\bf k},j,j')=\frac{3}{4}a_i(j,j')a_1(j)a_1(j'){\cal T}_{\bf k}(j,j')f^{(2)}_{\bf k}(j,j')\\ &&\quad-e_{i1}({\bf k},j-j')a_2(j,j')-e_{i2}({\bf k},j-j')a_1(j)\\
&&\quad-e_{i3}({\bf k},j-j')a_4(j,j')-e_{i4}({\bf k},j-j')a_3(j,j')\\
&&\quad+4p_i({\bf k},j-j')C'(j,j'),\\
&&b_i({\bf k},j,j')=-\frac{1}{4}a_i(j,j')a_1(j)a_1(j'){\cal T}_{\bf k}(j,j')f^{(1)}_{\bf k}(j,j')\\ &&\quad+c_{i1}({\bf k},j-j')\big[a_2(j,j')-T^{-1}\delta_{jj'}a_1(j)\big]+c_{i2}({\bf k},j-j')a_1(j)\\
&&\quad+c_{i3}({\bf k},j-j')a_4(j,j')+c_{i4}({\bf k},j-j')a_3(j,j').
\end{eqnarray*}

The above equations form a closed set allowing one to find the electron Green's function by iteration for given values of $U/t$, $T/t$, $\mu/t$, and functions $t_{\bf k}$, $v_{\bf k}$. The iteration procedure appears as follows: the initial or obtained in the previous step irreducible part $K({\bf k},j)$ is used for calculating the Green's function (\ref{Larkin}). The latter is applied for finding the renormalized hopping (\ref{theta}), renormalized interaction (\ref{phi}), and coefficients in the BSE equations (\ref{zi}) and (\ref{yi}). After their solution, we can calculate vertices (\ref{Vs2}) and (\ref{Va2}), from which the new function $K({\bf k},j)$ is derived from Eq.~(\ref{K}). The procedure is continued until convergence. As the starting function $K$ in this iteration, we used $C^{(1)}(j)$, the first term in this equation. It is the irreducible part of the Hubbard-I approximation \cite{Vladimir}. Investigating the order of the transition in the charge subsystem, we shall also use other iteration procedures, which will be discussed in the next section.

We must use a {\bf k} mesh corresponding to some finite cluster to perform momentum summations in the above formulas. With decreasing $T$, such a cluster goes into the saturated AFO and $\chi^{\rm sp}({\bf Q},0)\rightarrow\infty$ when $\xi$ approaches the cluster size. As a consequence, temperatures lower than the saturation temperature become unreachable. In Refs.~\cite{Sherman19a,Sherman21}, a cure was suggested for this difficulty. The magnetic saturation is connected with several terms in $K$ and $V^{(a)}$, which after summations over Matsubara frequencies, contain the multiplier $1/T$. We  substituted it with $1/(T+\zeta)$, where $\zeta$ was fitted such that $\chi^{\rm sp}({\bf Q},0)$ diverges at $T=0$, as is required by the Mermin-Wagner theorem for an infinite 2D crystal \cite{Mermin}. We used several momentum meshes from 8$\times$8 up to 32$\times$32 and compared several momentum sums in the above formulas and several local quantities. We found that these sums and quantities calculated with different meshes agree well, although $\zeta$ decreases several times when switching from the smallest to the largest cluster. This fact gives grounds to consider the obtained results for the local quantities as reasonable approximations for an infinite crystal and the use of the summation over a finite number of momenta as a method of approximate integration over the Brillouin zone. Below results are calculated using mainly an 8$\times$8 {\bf k} mesh, for $2t\leq U\leq8t$, $v\lesssim U/2$ and temperatures $0.1t\lesssim T\ll U$ to satisfy the condition (\ref{condition}) for $\mu=U/2$.

\section{Results}
\subsection{Determinants}
Phase transitions occur when the two-particle correlators $\widetilde{V}^{(s)}$ and $\widetilde{V}^{(a)}$ diverge or change discontinuously \cite{Abrikosov}. As follows from Eqs.~(\ref{Vts}) and (\ref{Vta}), the sources of such changes may be the renormalized Coulomb interaction $\varphi$ or vertices $V^{(s)}$ and $V^{(a)}$. Simple calculations show that the denominator in Eq.~(\ref{phi}) can vanish in the parameter range of interest. Indeed, let us approximate Green's function in this equation by the first-order cumulant (\ref{cumulants}). This approximation corresponds to the case $t\ll U$. Performing the summation over Matsubara frequencies in (\ref{phi}), we find that the denominator vanishes at $v=U/4$, $\nu=0$, and ${\bf k=Q}=(\pi/a,\pi/a)$ with $a=|{\bf a}|$. It is the result of the mean-field approximation \cite{Yan,Dagotto}. However, Green's functions obtained in our calculations differ significantly from $C^{(1)}$. We did not observe the divergence of $\varphi$ for any of the considered sets of parameters. All found divergencies and discontinuities were connected with the vertices $V^{(s)}$ and $V^{(a)}$.

The BSEs (\ref{Vs}) and (\ref{Va}) are linear systems of equations, and discontinuities in their solutions are defined by their determinants $\Delta_s({\bf k},\nu)$ and $\Delta_a({\bf k},\nu)$, for the stationary case at $\nu=0$. The value of ${\bf k}$, for which a discontinuity occurs, defines the character of the ordered state. In the 2D Hubbard model on an infinite lattice, $\Delta_a({\bf k},\nu)$ vanishes for $T\rightarrow0$ at ${\bf k=Q}$, $\nu=0$ and half-filling \cite{Sherman18}. This behavior signals the transition from the short-range to the long-range AFO. In this model, in the range of chemical potentials (\ref{condition}), $\Delta_s({\bf k},0)$ varies slowly near unity.

\begin{figure}[t]
\centerline{\resizebox{0.8\columnwidth}{!}{\includegraphics{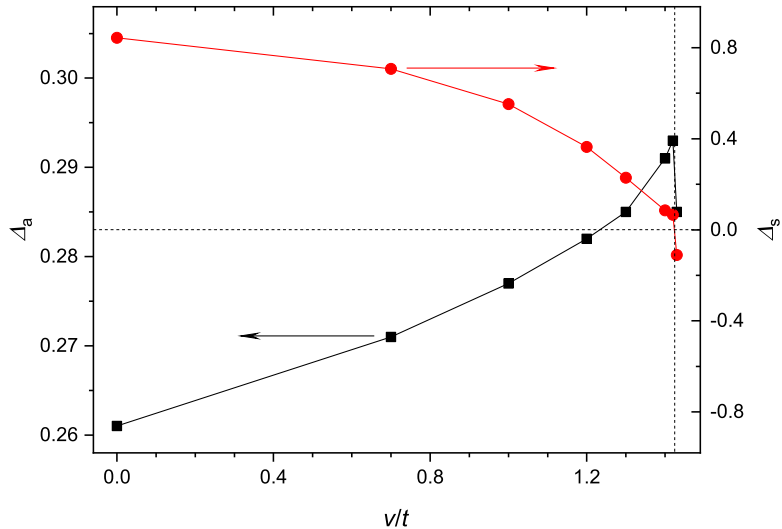}}}
\caption{Dependencies of the determinants $\Delta_s({\bf k},\nu)$ and $\Delta_a({\bf k},\nu)$ of the systems of linear equations (\ref{zi}) and (\ref{yi}) on the intersite interaction constant $v$ for ${\bf k=Q}$, $\nu=0$, $U=4t$, and $T=0.096t$. Red circles show calculated results for $\Delta_s$, the right coordinate axis. Black squares are data for $\Delta_a$, the left axis. The solid lines are a guide to the eye. The vertical dashed line indicates the location of the transition.}\label{Fig2}
\end{figure}
Above, we reduced the complex BSEs (\ref{Vs}) and (\ref{Va}) to two systems (\ref{zi}) and (\ref{yi}) containing every four equations. One can easily calculate their determinants. Results of such calculations for one of the parameter sets are shown in Fig.~\ref{Fig2}. The determinants for ${\bf k=Q}$ and $\nu=0$ are depicted since abrupt changes in $\Delta_s$ occur at these momentum and frequency. On the other hand, $\Delta_a({\bf Q},0)$ is the probe for the antiferromagnetic ordering since $\chi^{\rm sp}({\bf Q},0)\sim1/\Delta_a({\bf Q},0)$. As seen from the figure, $\Delta_s({\bf Q},0)$ abruptly changes sign at $v=v_c\approx1.425t$. We observed similar discontinuities of this determinant for other parameters in the range of the on-site repulsions $2t\leq U\lesssim5t$. Smaller values of $U$ were not considered. For larger on-site repulsions, we did not find such behavior of the denominator for $v\lesssim U/2$ and $T\ll U$.

This abrupt change in $\Delta_s({\bf Q},0)$ points to a phase transition in the charge subsystem. As will be seen below, the transition manifests itself in the susceptibility $\chi^{\rm ch}({\bf k},0)$, which peaks  sharply at ${\bf k=Q}$ for $v\approx v_c$ and changes sign at the transition. From earlier results, we know that this behavior is connected with the appearance of SAOs at the bottom of the electron spectrum.

\begin{figure}[t]
\centerline{\resizebox{0.8\columnwidth}{!}{\includegraphics{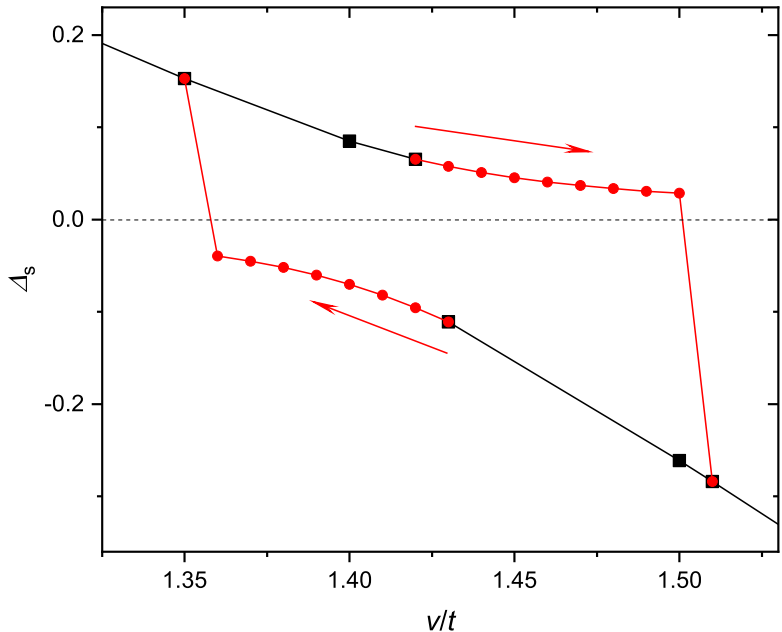}}}
\caption{The dependence of the determinant $\Delta_s({\bf Q},0)$ on $v$. Results obtained with the initial $K({\bf k},j)=C^{(1)}(j)$ are shown by black squares. Red circles correspond to solutions derived by a gradual variation of $v$ from the former results with $v$ closest to $v_c$. Arrows show the directions of these variations. $U=4t$, $T=0.096t$. The lines are a guide to the eye.}\label{Fig3}
\end{figure}
We noticed that the sign change in $\Delta_s$ occurs abruptly. Indeed, if we try to come close to a transition point, we obtain a solution with either positive or negative $\Delta_s({\bf Q},0)$ and never with a negligibly small value. This result points to the first-order transition. To prove this supposition, we obtained solutions in the transition region in a somewhat different manner. As indicated above, we mainly use iteration starting from $K({\bf k},j)=C^{(1)}(j)$. Now we use such obtained solutions, which are the closest to a transition point, as starting ones in iteration, in which we gradually vary $v$. This parameter is changed in the direction of the opposite side of the transition. That is, if, for example, we take a solution with $v>v_c$ as initial, this constant is slightly decreased. After achieving the convergence, the obtained $K$ is used as the starting one in the next iteration with an even smaller $v$. The determinants $\Delta_s({\bf Q},0)$ of such obtained solutions are shown in Fig.~\ref{Fig3}. As follows from the figure, in the range $1.36t\leq v\leq1.5t$, there are two coexisting solutions with opposite signs of the determinant. This coexistence is inherent in the first-order transitions. Similar behavior is also observed for other considered parameter sets with $U\lesssim5t$. The conclusion about the transition order agrees with the results of previous works \cite{Sun,Ayral,Loon,Paki,Aichhorn,Davoudi}.

As seen from Fig.~\ref{Fig3}, $\Delta_s$ have noticeably reduced with this new iteration scheme. Nevertheless, they remained nonzero, and their leveling at non-vanishing values is seen in the figure. Hence, the vertex $V^{(s)}({\bf Q},0)$ and the susceptibility $\chi^{\rm ch}({\bf Q},0)$ do not diverge at the transition. It means that the state at $v=v_c$ has short-range ordering. Similar behavior was observed for other considered sets of parameters. However, for higher temperatures, values of $|\Delta_s({\bf Q},0)|$ at $v\approx v_c$ appeared to be much smaller than for lower $T$. For example, for $U=4t$ and $T=0.58t$, this parameter is approximately four times smaller than at $T=0.096t$. Since the charge and spin fluctuations fall off with growing $T$, this result indicates that they are responsible for the short-range ordering of the state at $v=v_c$.

\begin{figure}[t]
\centerline{\resizebox{0.8\columnwidth}{!}{\includegraphics{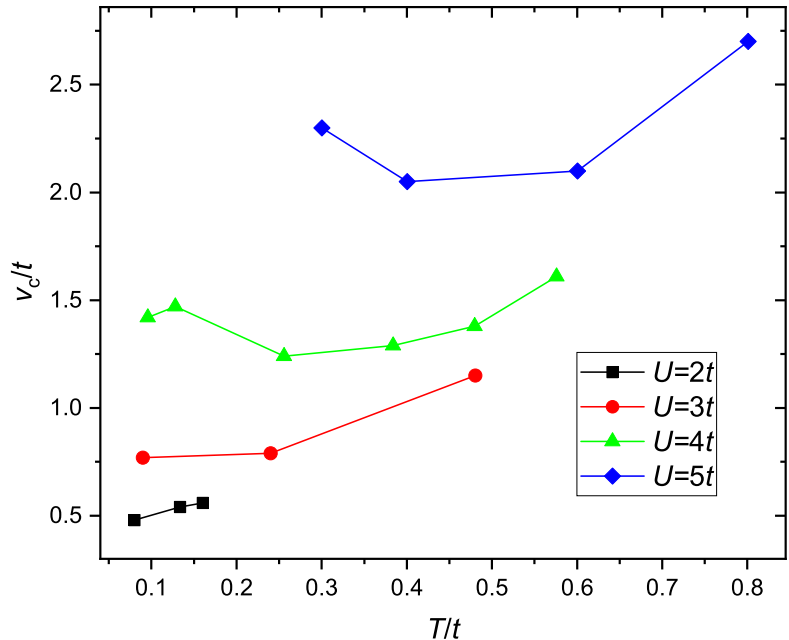}}}
\caption{The dependence of the critical value $v_c$ on the temperature $T$ and on-site repulsion $U$. Symbols show calculated results. Lines are a guide to the eye.}\label{Fig4}
\end{figure}
Figure~\ref{Fig4} demonstrates the dependence of the critical value $v_c$ on the temperature $T$ and on-site repulsion $U$. These values were obtained in iteration with the starting irreducible part $K({\bf k},j)=C^{(1)}(j)$. The figure shows $v$-$T$ phase diagrams for four values of $U$, in which the SAO region is located above the respective curve and metallic domains below it. For $U=2t$ and $3t$, temperatures lower than those shown in the figure were not considered. For $U=3t$, $4t$, and $5t$, temperatures higher than those shown in the figure were not used since they violate the condition (\ref{condition}). We found no phase transitions in the cases $U=2t$, $T\gtrsim0.16t$ and $U=5t$, $T\lesssim0.3t$ as well as for $U\gtrsim6t$ in the mentioned above ranges of $v$ and $T$. As the figure shows, $v_c$ depends rather strongly on $T$. For $U=2t$ and $3t$, the value $v_c$ is close to its mean-field estimate $U/4$ and exceeds it for larger $U$.

\begin{figure}[t]
\centerline{\resizebox{0.8\columnwidth}{!}{\includegraphics{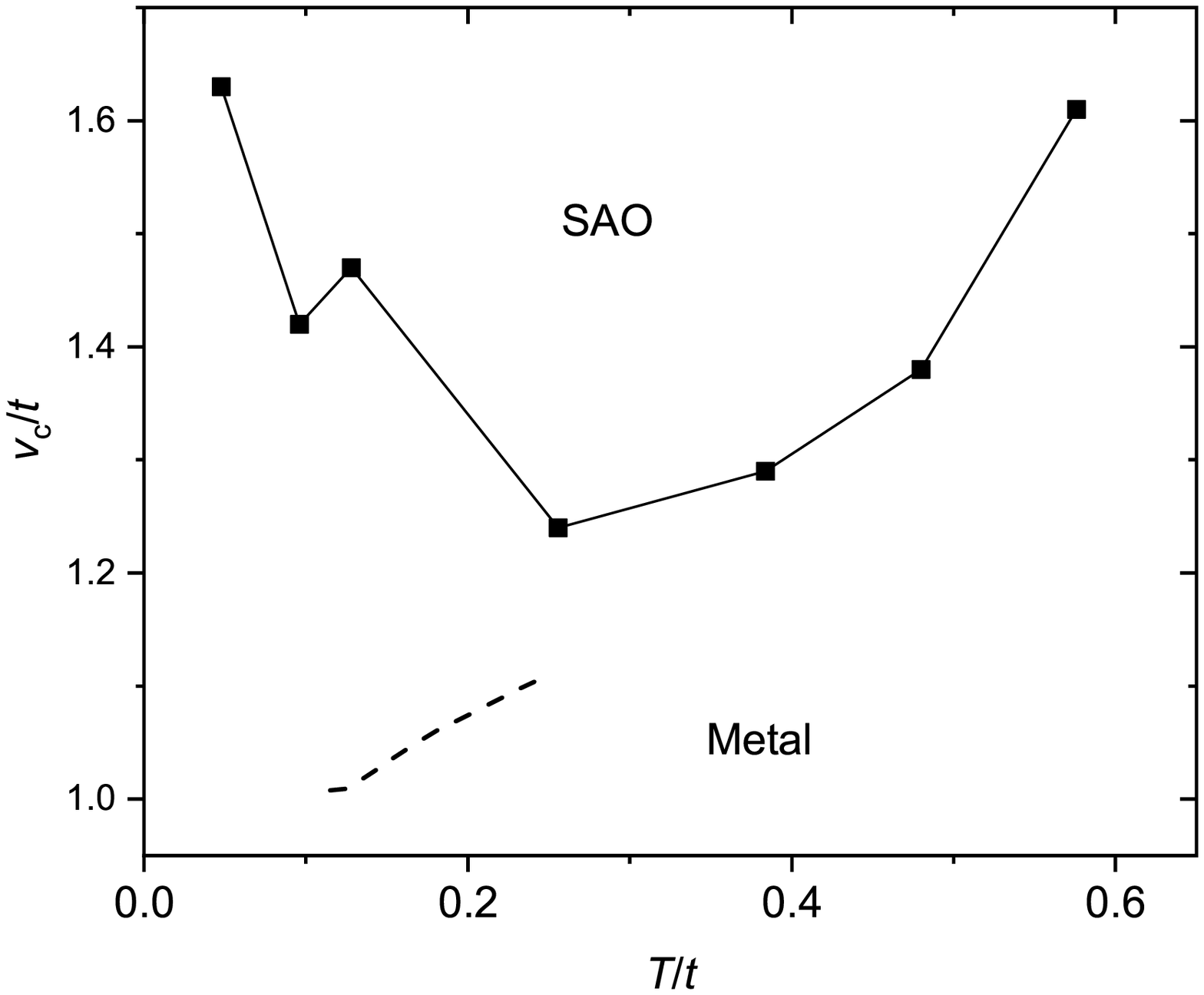}}}
\caption{The dependence of the critical value $v_c$ on the temperature $T$ for $U=4t$. Symbols show calculated results; lines are a guide to the eye. The dashed line shows the same dependence calculated using the dynamic cluster approximation with an 8-site cluster \protect\cite{Paki}.}\label{Fig5}
\end{figure}
In Fig.~\ref{Fig5}, the $v$-$T$ phase diagram of the crystal for $U=4t$ is shown on the broader temperature range. The solid curve is the same dependence $v_c(T)$, as depicted in Fig.~\ref{Fig4} for this value of $U$. As seen from Fig.~\ref{Fig5}, both with decreasing and increasing temperatures, larger and larger values of $v$ are necessary to stabilize SAOs. We suppose that the reasons for this behavior are thermal fluctuations for high temperatures and spin and charge fluctuations for low $T$. This influence of fluctuations explains why the curves for $U=2t$ and $5t$ in Fig.~\ref{Fig4} are bounded on one side. For $U=4t$, we did not find the transition to SAO for $T<0.04t$. One can suppose that dependencies $v_c(T)$ for all $U$ are bounded on low- and high-temperature sides.

We now focus on a kink near $T=0.1t$ on the curve in Fig.~\ref{Fig5}. The kink is located on the boundary between two distinct metallic states, one of which, on the low-temperature side, is characterized by the Slater dip at the Fermi level in the DOS, while the other by the narrow Fermi-level peak (see Fig.~\ref{Fig7} below). The DOS depression in the former state is connected with the Slater mechanism \cite{Slater} and a short-range AFO of electron spins. The Fermi-level peak in the latter state is a manifestation of the narrow band composed of the bound states of electrons and spin excitations \cite{Sherman20b,Sherman19}. By its nature, it is similar to the spin-polaron band of the $t$-$J$ model \cite{Schmitt,Ramsak,Sherman94}. The bound electron-spin-excitation states presume the existence of well-defined local spin moments. Hence the region of the phase diagram, characterized by the Fermi-level peak, have a higher degree of moment localization. Their existence is related to an increased spin entropy caused by the formation of local moments as the temperature grows \cite{Werner}. This behavior is analogous to the Pomeranchuk effect in liquid helium-3 \cite{Lee}. As seen in Fig.~\ref{Fig5}, the change of the moment localization manifests itself in the dependence $v_c(T)$.

For comparison, in Fig.~\ref{Fig5}, we reproduce the dependence $v_c(T)$ obtained using the dynamic cluster approximation with an 8-site cluster \cite{Paki}. Near $T=0.1t$, $v_c$ has the mean-field value $U/4=t$ -- charge and spin fluctuations taken into account in this approach do not reveal themselves here. For higher temperatures, $v_c$ grows presumably due to thermal fluctuations.

As mentioned, we observe the transition to SAOs in the range of on-site repulsions $U\lesssim5t$. This range is narrower than the interval $U\lesssim9t$ obtained for $v<U/2$ in the extended DMFT \cite{Ayral,Loon}. To elucidate the source of this difference, let us clarify the origin of the upper bound of $U$. The on-site and intersite repulsions compete with each other -- the former tends to the single site occupation, whereas the latter to alternating deviations from the such population on neighboring sites. Therefore, for $U>v$, the opening of the Mott gap at $U=U_c$ suppresses the phase transition in the charge subsystem. Hence, $U_c$ defines the range of on-site repulsions where this transition can be observed. For $T\approx 0.1t$, in the SCDT $U_c\approx 5.5t$ \cite{Sherman18}. The one-site DMFT gives the significantly overestimated \cite{Schafer} value $U_c\approx9t$, which explains the mentioned difference.

Let us return to Fig.~\ref{Fig2} and consider the dependence of $\Delta_a({\bf Q},0)$ on $v$. As mentioned above, this determinant is a probe of the AFO of electron spins in the system -- its zero value signals the establishment of the long-range order, while quantities $0<\Delta_a({\bf Q},0)$ point to a short-range ordering. The comparatively small $\Delta_a$ at $v=0$ corresponds to the antiferromagnetic correlation length $\xi\approx5a$. As $v_c$ is approached, the determinant grows, which points to a gradual decrease of $\xi$. The increase of $v$ leads to the transfer of SAOs to the lower part of the electron spectrum. In these states, occupation deviations from unity reduce site spins, which explains the decrease in $\xi$. The attenuation is not as strong as in small clusters \cite{Zhang,Paki}. The difference stems from the fact that SAOs destroy the saturated AFO of small lattices in the latter case, whereas our situation is far from the long-range ordering. The increase of $\Delta_a$ becomes more rapid as $v$ approaches $v_c$.

\subsection{Charge susceptibility}
\begin{figure*}[t]
\centerline{\resizebox{0.47\columnwidth}{!}{\includegraphics{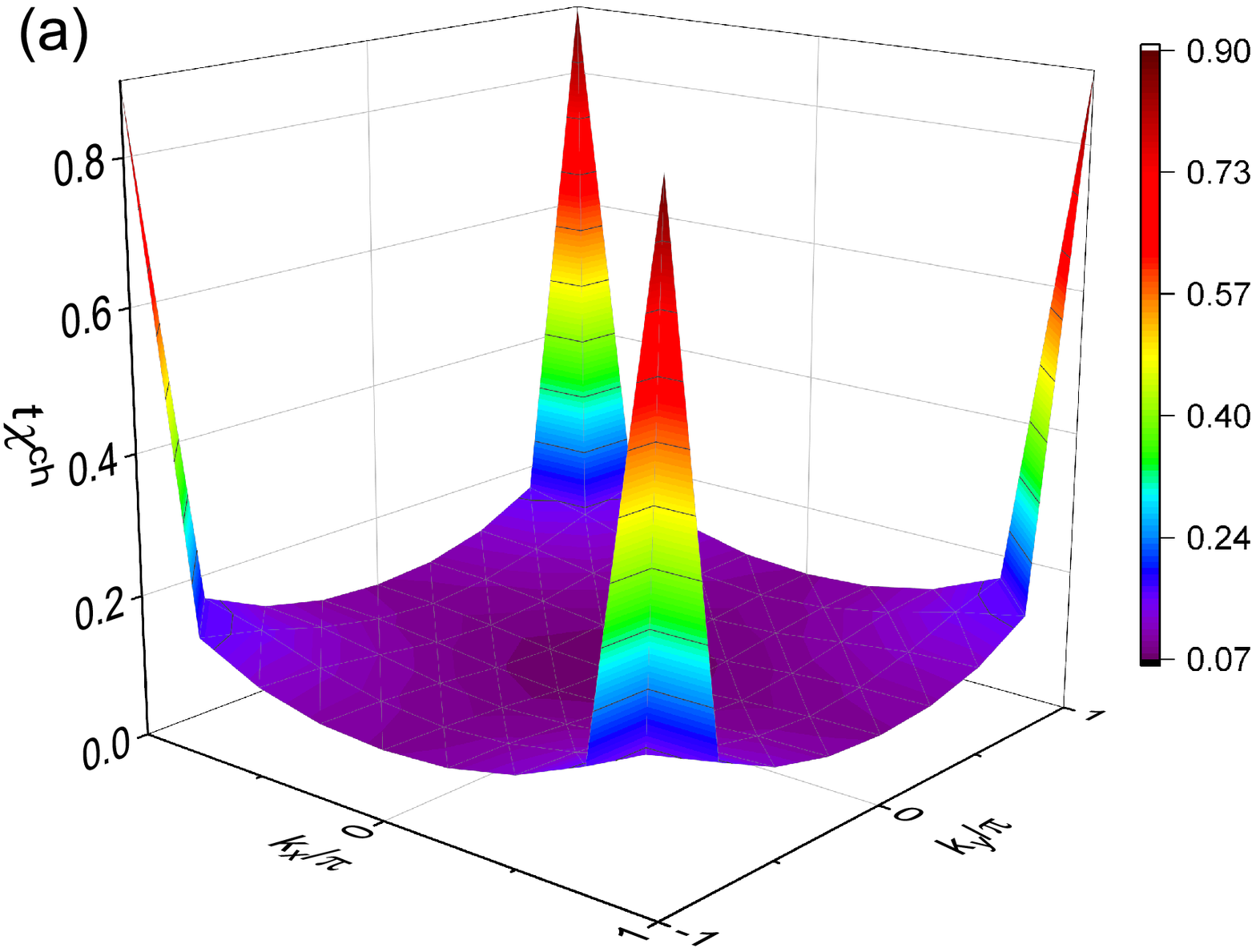}}\hspace{1.2em}\resizebox{0.47\columnwidth}{!}{\includegraphics{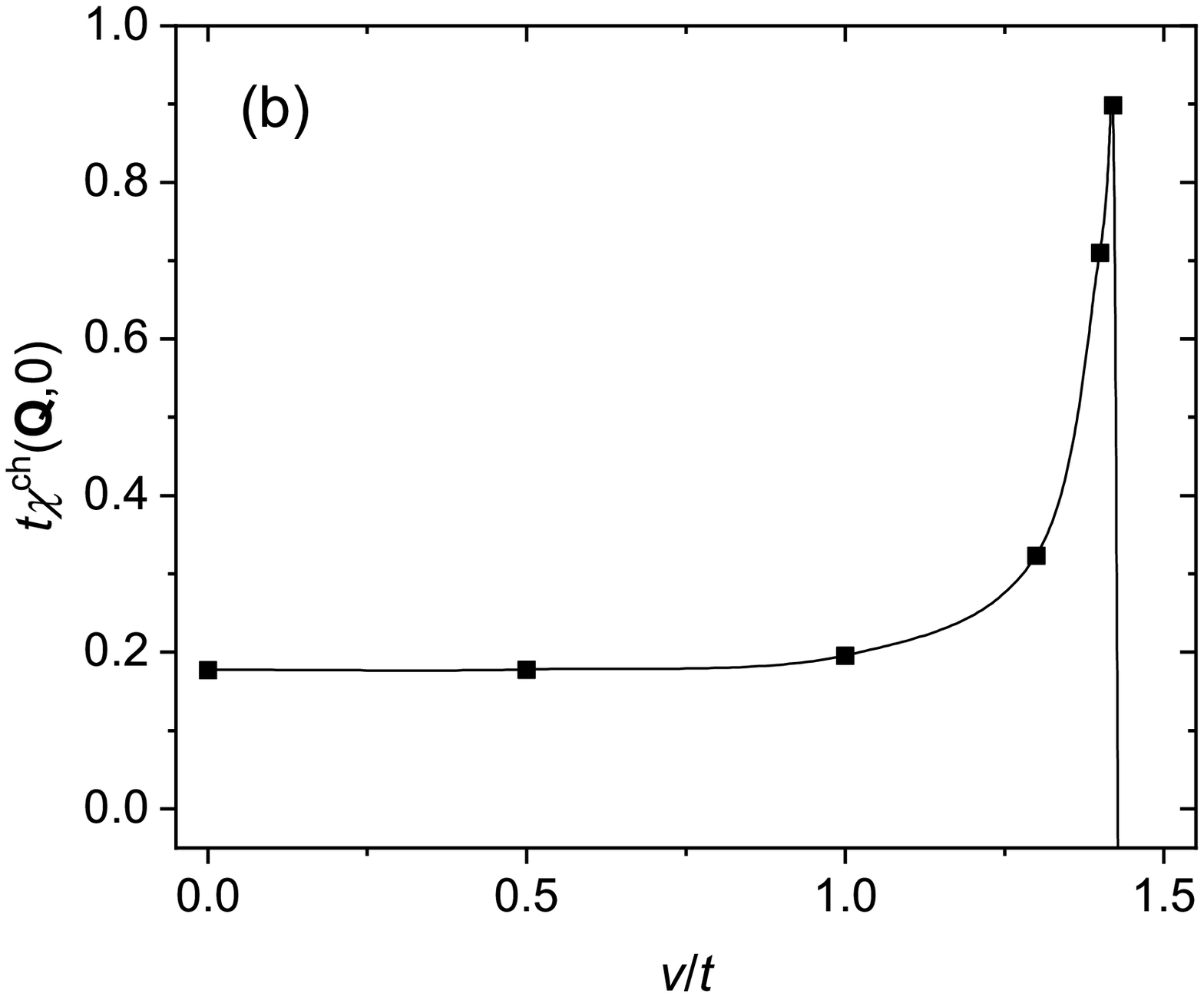}}}
\vspace{2ex}
\caption{The zero-frequency charge susceptibility $\chi^{\rm ch}({\bf k},0)$, Eq.~(\ref{chich}), as a function of the wave vector for $v=1.42t$ (a) and at ${\bf k=Q}$ as a function of $v$. $U=4t$, $T=0.096t$.}\label{Fig6}
\end{figure*}
The momentum dependence of the zero-frequency charge susceptibility $\chi^{\rm ch}({\bf k},0)$, Eq.~(\ref{chich}), calculated for $U=4t$, $T=0.096t$ and $v=1.42t$, in the nearest vicinity of the transition, is shown in Fig.~\ref{Fig6}(a). As the figure shows, the susceptibility peaks sharply at ${\bf k=Q}$. Its variation with $v$ at this momentum is depicted in Fig.~\ref{Fig6}(b). Near $v_c$, the susceptibility starts to grow rapidly and then sharply changes sign, which signals the phase transition in the charge subsystem. This dependence of $\chi^{\rm ch}({\bf k},0)$ on $v$ is connected with the behavior of the determinant $\Delta_s({\bf Q},0)$ discussed above. Analogous variations in the susceptibility are observed for other considered sets of parameters with $U\lesssim5t$.

The sharp peak of the zero frequency charge susceptibility
\begin{equation}\label{chiQ}
\chi^{\rm ch}({\bf k},0)=\frac{1}{2}\sum_{\bf l}\int_0^{1/T}{\rm e}^{{\rm i}{\bf kl}}\langle(n_{\bf l}(\tau)-\bar{n})(n_{\bf 0}-\bar{n})\rangle d\tau
\end{equation}
at ${\bf k=Q}$ and $v\approx v_c$ means the harmonic spatial variation of the occupation number correlator characterized by the momentum {\bf Q},
\begin{equation}\label{correlator}
\int_0^{1/T}\langle(n_{\bf l}(\tau)-\bar{n})(n_{\bf 0}-\bar{n})\rangle d\tau\sim{\rm e}^{{\rm i}{\bf Ql}}.
\end{equation}
Thus, occupation deviations from $\bar{n}=1$ have different signs on neighboring sites, indicating that SAOs are contained at the bottom of the electron spectrum.

As mentioned above, the susceptibility does not diverge at the transition point (see Fig.~\ref{Fig6}(b)). Finite values of $\chi^{\rm ch}({\bf Q},0)$ were related to charge and spin fluctuations taken into account in this work. As a consequence, the correlator (\ref{correlator}) decays with growing $|{\bf l}|$, and the decay coefficient is determined by the width of the maximum in the susceptibility. Hence the state at $v=v_c$ has a short-range ordering of alternating deviations of electron occupations.

\subsection{Density of states}
In this subsection, we consider the DOS of the obtained solutions,
\begin{equation*}
\rho(\omega)=-\frac{1}{\pi N}\sum_{\bf k}{\rm Im}G({\bf k},\omega).
\end{equation*}
The analytic continuation to real frequencies $\omega$ was performed using the maximum entropy method \cite{Press,Jarrell,Habershon}.

\begin{figure}[t]
\centerline{\resizebox{0.47\columnwidth}{!}{\includegraphics{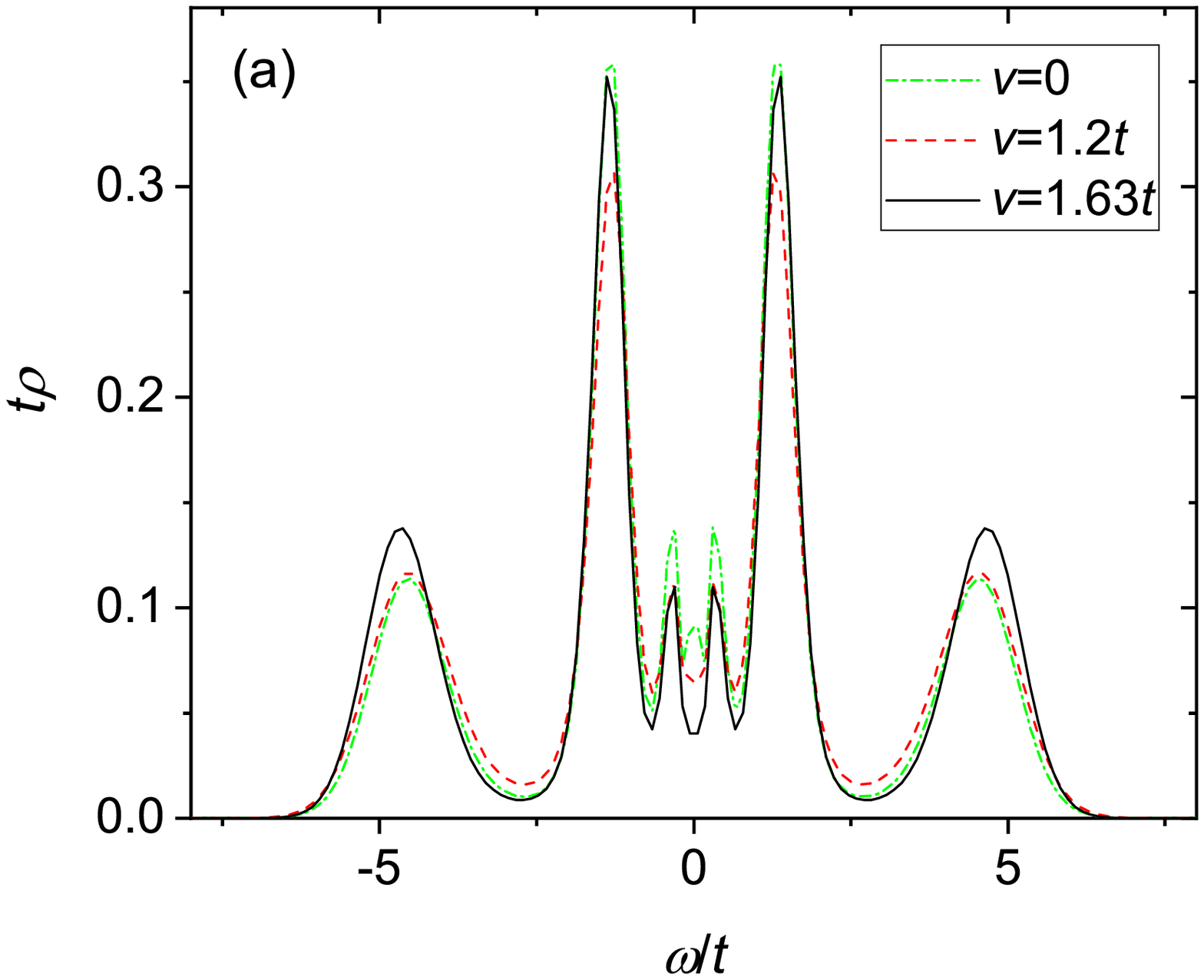}}\hspace{1.2em}\resizebox{0.47\columnwidth}{!}{\includegraphics{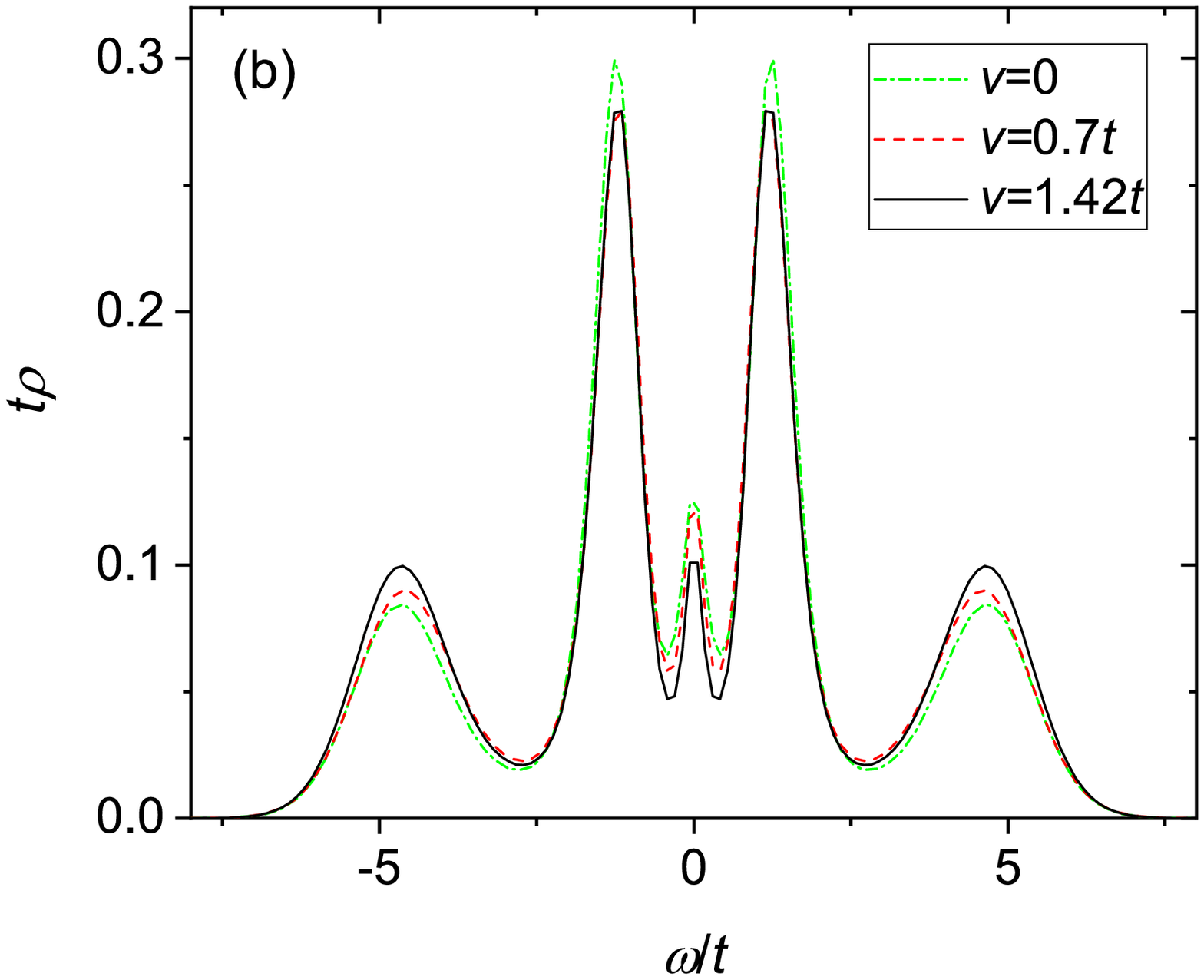}}}
\caption{The evolution of the density of states with the variation of $v$. $U=4t$, $T=0.048t$ (a) and $0.096t$ (b), values of $v$ are shown in the legends.}\label{Fig7}
\end{figure}
The evolution of the DOS with the variation of $v$ from zero up to $v\approx v_c$ is shown in Fig.~\ref{Fig7}. As mentioned above, for $v<U/2$, the region of SAO borders with metallic domains, two of which are characterized by the Slater dip and Fermi-level peak. Their DOSs are shown in Fig.~\ref{Fig7}. Mechanisms leading to the dip and the peak at the Fermi level are discussed in Subsection 3.1. From Fig.~\ref{Fig7}, we see these spectral peculiarities are retained when $v\rightarrow v_c$. However, central parts of spectra lose intensity with increasing $v$. It can be expected because, for $v>v_c$, a gap caused by the intersite repulsion is presumed to appear at the Fermi level.

\begin{figure}[t]
\centerline{\resizebox{0.8\columnwidth}{!}{\includegraphics{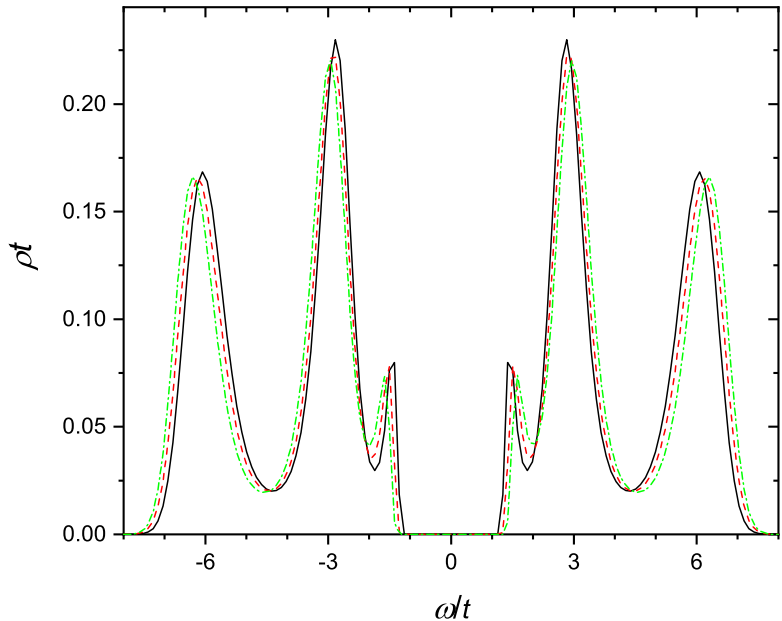}}}
\caption{Same as in Fig.~\ref{Fig7}, but for $U=8t$, $T=0.13t$. The black solid, red dashed, and green dash-dotted lines correspond to $v=0$, $1.78t$, and $3t$, respectively.}\label{Fig8}
\end{figure}

As mentioned above, in the considered ranges of parameters, we found no phase transitions and manifestations of SAOs for $U\gtrsim6t$. Respectively, the influence of the intersite repulsion on spectra is weaker in this case. An example of such changes is shown in Fig.~\ref{Fig8} for $U=8t$, $T=0.13t$. The main spectral modification caused by increasing $v$ is a tiny growth of the Mott gap with minor variations in DOS shapes. One can expect the gain in the gap width as $v$ grows from the fact that for $v\gg U,t$, the system is an insulator with the gap $\sim 8v$. Indeed, in this case, the lowest states are formed from nearly empty and doubly occupied site states. The transfer of an electron from a doubly occupied to an empty site needs the energy input of $8v$. Figure~\ref{Fig8} shows that the gap increases starting from small $v$. This result contradicts the data obtained with the extended DMFT. In the latter approach, the gap decreases for $v\lesssim U$ \cite{Ayral}. In the cluster DMFT \cite{Paki}, the gap grows with $v$ as in our results in Fig.~\ref{Fig8}. However, in this approach, the small used cluster is in the saturated antiferromagnetic state for moderate temperatures. Consequently, even for moderate $U$, the Slater gap \cite{Slater} is observed for small $v$ instead of the Fermi-level peak. The gap width grows with $v$ and transforms gradually to the intersite-repulsion gap.

\section{Conclusion}
In this work, we used the strong coupling diagram technique for investigating the extended Hubbard model at half-filling. This approach applies the series expansion over the kinetic and intersite repulsion terms for calculating Green's functions. We summed infinite series of diagrams for a two-dimensional square lattice. It allowed us to properly account for spin and charge fluctuations and actual short-range antiferromagnetic ordering in the crystal for finite temperatures. The ranges of the on-site Coulomb repulsion $2t\leq U\leq8t$, intersite interaction $0\leq v\lesssim U/2$, and temperature $0.1t\lesssim T\ll U$ were considered. Here $t$ is the hopping constant between neighboring sites. We found that for $U\lesssim5t$ the zero-frequency charge susceptibility at the corner of the Brillouin zone $\chi^{\rm ch}({\bf Q},0)$ abruptly changes sign at $v=v_c\gtrsim U/4$. The susceptibility does not diverge at this value of the intersite repulsion; however, it peaks sharply at the momentum ${\bf k=Q}$. It indicates that states with alternating deviations from the mean occupation on neighboring sites comprise the bottom of the electron spectrum for $v$ near $v_c$. The transition is of the first order -- two solutions coexist neat the transition point. As follows from the temperature dependence of determinants of the Bethe-Salpeter equation, finite values of the charge susceptibility at $v=v_c$ are a consequence of charge and spin fluctuations taken into account in our approach. Hence the state at this $v$ has a short-range ordering of alternating populations, and the width of the maximum in $\chi^{\rm ch}({\bf k},0)$ at ${\bf k=Q}$ defines its correlation length. This result differs from works using mean-field approximations, in which the susceptibility diverges. The growth of $v$ leads to the decay of antiferromagnetic spin correlations. For $U\lesssim5t$, peculiarities of the metallic densities of states -- the Slater dip and Fermi-level peak -- are retained up to $v_c$. However, the central parts of the spectra lose intensity. This result may indicate a gap at the Fermi level for $v>v_c$. In the insulating cases $U\gtrsim6t$ and $T\approx0.1t$, for which we found no transitions in the charge subsystem in the considered range of parameters, a monotonous growth of the Mott gap with increasing $v$ is observed.

\end{document}